\documentclass[10pt,a4paper]{article}
\usepackage{amsmath, amsfonts, amssymb}
\usepackage{graphicx}
\usepackage{url}
\usepackage{natbib}

\begin{document}
\title{\bf \large Analyzing Astronomical Data with Machine Learning Techniques }
\author{Mohammad H. Zhoolideh Haghighi\thanks{E-mail: zhoolideh@kntu.ac.ir}
}
\date{\it  \small   Department of Physics, K.N. Toosi University of Technology, Tehran P.O. Box 15875-4416, Iran, \\
School of Astronomy, Institute for Research in Fundamental Sciences (IPM), Tehran, 19395-5746, Iran,
}

\maketitle

\begin{abstract}
Classification is a popular task in the field of Machine Learning (ML) and Artificial Intelligence (AI), and it happens when outputs are categorical variables. There are a wide variety of models that attempts to draw some conclusions from observed values, so classification algorithms predict categorical class labels and uses it in classifying new data.
Popular classification models including logistic regression, decision tree, random forest, Support Vector Machine (SVM), multilayer perceptron, naive bayes, neural networks have proven to be efficient and accurate applied on many industrial and scientific problems. Particularly, application of ML to astronomy has shown to be very useful for  classification, clustering and data cleaning. It is because after learning computers, these tasks can be done automatically by them in a more precise and  more rapid way than human operators. In view of this, in this paper, we will review some of these popular classification algorithms, and then we apply some of them to the observational data of nonvariable and the RR Lyrae variable stars that come from the SDSS survey. For the sake of comparison, we calculate the accuracy and $F_1$-score of the applied models.

\end{abstract}


\section{\large Introduction}

Machine learning is a branch of artificial intelligence (AI) and computer science which focuses on the use of data and algorithms to imitate the way that humans learn. One of the most popular branches of ML is Supervised Learning \citep{Jain}. It is a machine learning approach defined by its use of labeled datasets to train algorithms to classify data and predict outcomes. The labeled dataset, which is called training set, has output labeled corresponding to input data. The machine is supposed to understand what to search for, in the unseen data and provide some predictions based on what it has already learned from the labeled data.
 There are two main areas where supervised machine learning comes in handy, classification problems and regression problems. Classification refers to taking an input value and mapping it to a discrete value. In classification problems, our output typically consists of classes or categories, but regression is related to continuous data and in that the predicted output values are real numbers \citep{Dutton}.

As astronomy is experiencing a rapid growth in the size and complexity of data, these changes foster the development of data-driven science as a useful companion to the common model-driven data analysis paradigm, where astronomers develop automatic tools to mine datasets and extract novel information from them. In recent years, machine learning algorithms have become increasingly popular among astronomers, and ML is now used for a wide variety of tasks. Machine learning algorithms can be used in the field of astronomy and cosmology for astronomical object detection and object classification \citep{Gordon}, clustering \citep{Campana}, anomaly detection \citep{Anomaly}, feature extraction \citep{Sen} and other related domains.

\section{\large Review; Popular Classification Algorithms }

\subsection{Linear regression}
Simple linear regression is a type of regression analysis where there is one independent  variable (x) with  a linear relationship with another dependent variable (y). Based on the given data points, the algorithm try to find a line that models the points in the best possible way. The line can be modeled based on a linear equation \citep{Efron}
 
 \begin{equation}
 y = a * x + b
 \end{equation}
 The motivation of the linear regression algorithm is to find the best values for $a$ and $b$.
 The cost function helps us to figure out the best possible values for $a$ and $b$ which would provide the best fit line for the data points. Since the best values for $a$ and $b$ are desired, we convert this search problem into a minimization problem where we would like to minimize the error between the predicted value and the actual value.
 
 \begin{equation}
 \operatorname{Minimize} \frac{1}{n} \sum_{i=1}^{n}\left(  y_{pred}-y_{i}\right)^{2}
 \end{equation}

\begin{equation}
J=\frac{1}{n} \sum_{i=1}^{n}\left( y_{pred}-y_{i}\right)^{2}
 \end{equation}
We choose the above function for minimization. The difference between the predicted values and ground truth measures the error difference. We square the error difference and sum over all data points and divide that value by the total number of data points. This provides the average squared error over all the data points. Now, using this cost function we can change the values of $a$ and $b$ such that the cost function value settles at the minima. To minimize the cost function, one can use gradient descent which is a method of updating $a$ and $b$ to reduce the cost function. The idea is that we start with some values for $a$ and $b$ and then we change these values iteratively to reduce the cost. Gradient descent helps us on how to change the values.
To update $a$ and $b$, we take gradients from the cost function. To find these gradients, we take partial derivatives with respect to $a$ and $b$ as follows: 

\begin{equation}
\begin{gathered}
J=\frac{1}{n} \sum_{i=1}^{n}\left(y_{pred}-y_{i}\right)^{2} \\
J=\frac{1}{n} \sum_{i=1}^{n}\left(a+b \cdot x_{i}-y_{i}\right)^{2} \\
\frac{\partial J}{\partial a}=\frac{2}{n} \sum_{i=1}^{n}\left(a+b \cdot x_{i}-y_{i}\right) \Longrightarrow \frac{\partial J}{\partial a}=\frac{2}{n} \sum_{i=1}^{n}\left(y_{pred}-y_{i}\right) \\
\frac{\partial J}{\partial b}=\frac{2}{n} \sum_{i=1}^{n}\left(a+b \cdot x_{i}-y_{i}\right) \cdot x_{i} \Longrightarrow \frac{\partial J}{\partial b}=\frac{2}{n} \sum_{i=1}^{n}\left(y_{pred}-y_{i}\right) \cdot x_{i}
\end{gathered}
\end{equation}

\begin{equation}
\begin{gathered}
a=a-\alpha \cdot \frac{2}{n} \sum_{i=1}^{n}\left(y_{pred}-y_{i}\right) \\
b=b-\alpha \cdot \frac{2}{n} \sum_{i=1}^{n}\left(y_{pred}-y_{i}\right) \cdot x_{i}
\end{gathered}
\end{equation}
The partial derivates are the gradients and they are used to update the values of $a$ and $b$. $\alpha$ is the learning rate which is a hyper-parameter that must be specified. A smaller learning rate could help to be closer to the minima but takes more time to reach the minima, a larger learning rate converges sooner but there is a chance that one could overshoot the minima.
 
\subsection{Logistic regression}
Logistic regression is another powerful supervised ML algorithm used for binary classification problems (when target is categorical). The best way to think about logistic regression is that it is a linear regression but for classification problems. Logistic regression essentially uses a logistic function defined below to model a binary output variable \citep{Sur}. The primary difference between linear regression and logistic regression is that logistic regression's range is bounded between 0 and 1. In addition, as opposed to linear regression, logistic regression does not require a linear relationship between inputs and output variables.
In this approach, instead of defining cost function based on $y$, the sigmoid of $y$ is used. 

\begin{equation}
 \begin{array}{l}
Output =0 ~ or ~ 1 \\ \\
y =a * x+ b  \\ \\
\mathrm{h} \Theta(\mathrm{x})=\operatorname{sigmoid}(\mathrm{y}) \\ \\
\operatorname{sigmoid}(y)=\frac{1}{1+e^{-y}}\\ \\

  \end{array}
\end{equation}

\begin{equation}
\begin{aligned}
\operatorname{Cost}(h \ominus(x), y(\text { actual }))=&-\log (h \ominus(x)) \text { if } y=1 \\
&-\log (1-h \ominus(x)) \text { if } y=0
\label{costlog}
\end{aligned}
\end{equation}
By having the cost function of logistic regression as defined in equation (\ref{costlog}), one can easily find $a$ and $b$ via minimizing this cost function; as a result $y$ will be determined.

\subsection{Naive Bayesian (NB)}
Using Bayes theorem, we can find the probability of A, happening, given that B has occurred. Here, B is the evidence and A is the hypothesis. The assumption made here is that the predictors/features are independent. Bayes theorem can be rewritten as:
 
 \begin{equation}
 P(y \mid X)=\frac{P(X \mid y) P(y)}{P(X)}
\end{equation}
Where X is the feature vector and by substituting for X and expanding using the chain rule we get,

\begin{equation}
P\left(y \mid x_{1}, \ldots, x_{n}\right)=\frac{P\left(x_{1} \mid y\right) P\left(x_{2} \mid y\right) \ldots P\left(x_{n} \mid y\right) P(y)}{P\left(x_{1}\right) P\left(x_{2}\right) \ldots P\left(x_{n}\right)}
\end{equation}
Now, you can obtain the values for each by looking at the dataset and substitute them into the equation. For all entries in the dataset, the denominator does not change, it remain static. Therefore, the denominator can be removed and a proportionality can be introduced.
 \begin{equation}
 P(y \mid x_1,..., x_n) \propto P( y) ~\Pi_i^n~P(x_i\mid y)
\end{equation}
One popular type of Naive Bayes Classifier is Gaussian Naive Bayes. When the predictors take up a continuous value and are not discrete, we assume that these values are sampled from a gaussian distribution. Since the way the values are present in the dataset changes, the formula for conditional probability changes to,

\begin{equation}
P\left(x_{i} \mid y\right)=\frac{1}{\sqrt{2 \pi \sigma_{y}^{2}}} \exp \left(-\frac{\left(x_{i}-\mu_{y}\right)^{2}}{2 \sigma_{y}^{2}}\right)
\end{equation}
In our case, the class variable(y) has only two outcomes, yes or no. There could be cases where the classification could be multivariate. Therefore, we need to find the class y with maximum probability.

 \begin{equation}
 y = \text{argmax} P( y) ~\Pi_i^n~P(x_i\mid y)
\end{equation}
 Naive Bayes algorithms are mostly used in sentiment analysis, spam filtering, recommendation systems etc. They are fast and easy to implement but their biggest disadvantage is that the requirement of predictors to be independent \citep{Nb}. In most of the real cases, the predictors are dependent, this hinders the performance of the classifier.

\subsection{Support Vector Machines (SVMs)}
Support vector machine is highly preferred by many as it produces significant accuracy with less computation power. Support Vector Machine, abbreviated as SVM can be used for both regression and classification tasks. But, it is widely used in classification objectives. The objective of the support vector machine algorithm is to find a hyperplane in an N-dimensional space(N — the number of features) that distinctly classifies the data points \citep{SVM}. To separate the two classes of data points, there are many possible hyperplanes that could be chosen. Our objective is to find a plane that has the maximum margin, i.e the maximum distance between data points of both classes. Maximizing the margin distance provides some reinforcement so that future data points can be classified with more confidence. Hyperplanes are decision boundaries that help classify the data points. Data points falling on either side of the hyperplane can be attributed to different classes. Also, the dimension of the hyperplane depends upon the number of features. In logistic regression, we take the output of the linear function and squash the value within the range of [0,1] using the sigmoid function. If the squashed value is greater than a threshold value(0.5) we assign it a label 1, else we assign it a label 0. In SVM, we take the output of the linear function and if that output is greater than 1, we identify it with one class and if the output is -1, we identify is with another class. Since the threshold values are changed to 1 and -1 in SVM, we obtain this reinforcement range of values([-1,1]) which acts as margin. 
 
 In the SVM algorithm, we are looking to maximize the margin between the data points and the hyperplane. The loss function that helps maximize the margin is Hinge loss. The cost is 0 if the predicted value and the actual value are of the same sign. If they are not, we then calculate the loss value. $c(x, y, f(x))=\left\{\begin{array}{ll}0, ~~  \text { if } y * f(x)  \geq 1 &\\ 1-y * f(x), \quad  \text { else }\end{array} \quad \right.$ 
 
 We also add a regularization parameter the cost function. The objective of the regularization parameter is to balance the margin maximization and loss. After adding the regularization parameter, the cost functions looks as below.

\begin{equation}
\min _{w} [\lambda\|w\|^{2}+\sum_{i=1}^{n}\left(1-y_{i}\left\langle x_{i}, w\right\rangle\right)_{+}]
 \end{equation}
 
 By taking partial derivatives with respect to the weights to find the gradients. Using the gradients, we can update our weights.
 
 \begin{equation}
 \frac{\delta}{\delta w_{k}} \lambda\|w\|^{2}=2 \lambda w_{k}
 \end{equation}
 
 \begin{equation}
 \frac{\delta}{\delta w_{k}}\left(1-y_{i}\left\langle x_{i}, w\right\rangle\right)_{+}= \begin{cases}0, & \text { if } y_{i}\left\langle x_{i}, w\right\rangle \geq 1 \\ -y_{i} x_{i k}, & \text { else }\end{cases}
 \end{equation}
 When there is no misclassification, i.e our model correctly predicts the class of our data point, we only have to update the gradient from the regularization parameter.
 
 \begin{equation}
 w=w-\alpha \cdot(2 \lambda w)
\end{equation}
When there is a misclassification, i.e our model make a mistake on the prediction of the class of our data point, we include the loss along with the regularization parameter to perform gradient update.
\begin{equation}
 w=w+\alpha \cdot(y_{i} \cdot x_{i }-2 \lambda w)
\end{equation}

\subsection{K-means}
K-means is an unsupervised classification algorithm, also called clusterization, that groups objects into k groups based on their characteristics. The grouping is done by minimizing the sum of the distances between each object and the group or cluster centroid \citep{Kmeans}. The distance that is usually used is the quadratic or euclidean distance. The algorithm has three steps:
1) Initialization: once the number of groups, k has been chosen, k centroids are established in the data space, for instance, choosing them randomly, 2) Assignment of objects to the centroids: each object of the data is assigned to its nearest centroid, 3) Centroids update: The position of the centroid of each group is updated taking as the new centroid the average position of the objects belonging to said group.
 Steps 2 and 3 should be repeated until the centroids do not move, or move below a threshold distance in each step.
The k-means algorithm solves an optimization problem and the function to be optimized (minimized) is the sum of the quadratic distances from each object to its cluster centroid.
The objects are represented with d dimension vectors $\left(\mathbf{x}_{1},\mathbf{x}_{2},\ldots,\mathbf{x}_{n}\right)$ and the algorithm k-means builds k groups where the sum of the distances of the objects to its centroid is minimized within each group $\mathbf{S}=\left\{ S_{1},S_{2},\ldots,S_{k}\right\}$. The problem can be formulated as:

\begin{equation}
\underset{\mathbf{S}}{\mathrm{min}}\; E\left(\boldsymbol{\mu_{i}}\right)=\underset{\mathbf{S}}{\mathrm{min}}\sum_{i=1}^{k}\sum_{\mathbf{x}_{j}\in S_i}\left\Vert \mathbf{x}_{j}-\boldsymbol{\mu}_{i}\right\Vert ^{2}
\end{equation}
where S is the dataset whose elements are the objects $x_j$ represented by vectors, where each of its elements represents a characteristic or attribute. We will have k groups or clusters with their corresponding centroid $\boldsymbol{\mu_{i}}$.
In each centroid update, from the mathematical point of view, we impose the extreme (minimum, in this case) necessary condition to the function $E\left(\boldsymbol{\mu_{i}}\right)$
that, for this quadratic function (1) is

\begin{equation}
\frac{\partial E}{\partial\boldsymbol{\mu}_{i}}=0\;\Longrightarrow\;\boldsymbol{\mu}_{i}^{(t+1)}=\frac{1}{\left|S_{i}^{(t)}\right|}\sum_{\mathbf{x}_{j}\in S_{i}^{(t)}}\mathbf{x}_{j}
\end{equation}
and the solution is to take each group element average as a new centroid. We have used the gradient descent method. The main advantages of the k-means method are that it is simple and fast. But it is necessary to decide the value of k and the final result depends on the initialization of the centroids. Also, it does not necessarily converge to the global minimum but to a local minimum.

\subsection{Decision Trees}
Decision Trees are trees that classify objects by sorting them based on feature values. Each node in a decision tree represents a feature in an instance to be classified, and each branch represents a value that the node can assume. Objects are classified starting at the root node and sorted based on their feature values. As the name goes, it uses a tree-like model of decisions. In the first split or the root, all attributes/features are considered and the training data is divided into groups based on this split \citep{DT}. Suppose we have three features, so we will have three candidate splits. So one can calculate how much accuracy each split will cost, using a function. The split that costs least is chosen.
When there is a large set of features, it results in large number of split, which in turn gives a huge tree. Such trees are complex and can lead to overfitting. So, we need to know when to stop. One way of doing this is to set maximum depth of your model. Maximum depth refers to the the length of the longest path from a root to a leaf.

Decision Trees are a type of Supervised Machine Learning (that is you explain what the input is and what the corresponding output is in the training data) where the data is continuously split according to a certain parameter. The tree can be explained by two entities, namely decision nodes and leaves. The leaves are the decisions or the final outcomes. And the decision nodes are where the data is split. There are many algorithms out there which construct Decision Trees, but one of the best is called as ID3 Algorithm. ID3 Stands for Iterative Dichotomiser 3. Before discussing the ID3 algorithm, we’ll go through few definitions.

Entropy, also called as Shannon Entropy is denoted by $H(S)$ for a finite set S, is the measure of the amount of uncertainty or randomness in data. 

\begin{equation}
H(S)=\sum_{x \in X} p(x) \log _{2} \frac{1}{p(x)}
\end{equation}

Decision Trees modified Intuitively, it tells us about the predictability of a certain event. Example, consider a coin toss whose probability of heads is 0.5 and probability of tails is 0.5. Here the entropy is the highest possible, since there’s no way of determining what the outcome might be. Alternatively, consider a coin which has heads on both the sides, the entropy of such an event can be predicted perfectly since we know beforehand that it’ll always be heads. In other words, this event has no randomness hence it’s entropy is zero. In particular, lower values imply less uncertainty while higher values imply high uncertainty.

Information gain is also called as Kullback-Leibler divergence denoted by IG(S,A) for a set S is the effective change in entropy after deciding on a particular attribute A. It measures the relative change in entropy with respect to the independent variables $I G(S, A)=H(S)-H(S, A)$.  Alternatively, 

\begin{equation}
I G(S, A)=H(S)-\sum_{i=0}^{n} P(x) * H(x)
\end{equation}
Where $IG(S, A)$ is the information gain by applying feature A. $H(S)$ is the Entropy of the entire set, while the second term calculates the Entropy after applying the feature A, where $P(x)$ is the probability of event x.

ID3 Algorithm will perform following tasks recursively. a) Create root node for the tree
, b) If all examples are positive, return leaf node ‘positive’
, c) Else if all examples are negative, return leaf node ‘negative’
, d) Calculate the entropy of current state H(S)
, e) For each attribute, calculate the entropy with respect to the attribute ‘x’ denoted by H(S, x)
, g) Select the attribute which has maximum value of IG(S, x)
, h) Remove the attribute that offers highest IG from the set of attributes
, i) Repeat until we run out of all attributes, or the decision tree has all leaf nodes.

\subsection{Neural Networks}
\begin{figure}[th!]
\begin{center}
\includegraphics[width=0.95\textwidth]{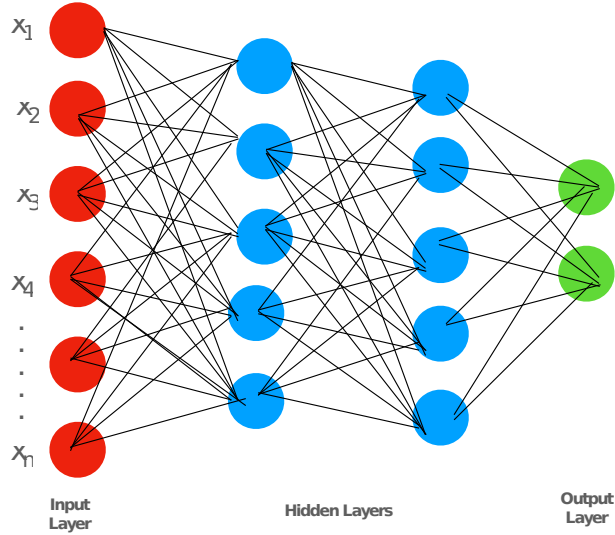}
\end{center}
\caption{A schematic view of a Neural Network. As can be seen, there are three main parts, input layer, hidden layers, and output layers.}
 \label{Fig0}
\end{figure}

Neural networks are multi-layer networks of neurons that are used to classify things and make predictions.  A neural network is made up of densely connected processing nodes \citep{NN}, similar to neurons in the brain. Each node may be connected to different nodes in multiple layers above and below it. These nodes move data through the network in a feed-forward fashion, meaning the data moves in only one direction. The node “fires” like a neuron when it passes information to the next node. A simple neural network has an input layer, output layer and one hidden layer between them. A network with more than three layers, including the input and output, is known as a deep learning network. In a deep learning network, each layer of nodes trains on data based on the output from the previous layer. Having more layers leads to, the greater the ability to recognize more complex information, based on data from the previous layers. The network makes decisions by assigning each connected node to a number known as a “weight.” The weight represents the value of information assigned to an individual node (i.e., how helpful it is in correctly classifying information). When a node receives information from other nodes, it calculates the total weight or value of the information. If the number exceeds a certain threshold, the information is passed onto the next layer. If the weight is below the threshold, the information is not passed on. In a newly formed neural network, all weights and thresholds are set to random numbers. As training data is fed into the input layer, the weights and thresholds refine to consistently yield correct outputs. In Figure (\ref{Fig0}), one can see the different parts of a Neural Network.

\section{\large Data}

Here we are going to introduce a real sample of astronomical data. RR Lyrae variables are periodic variable stars, commonly found in globular clusters \citep{Lyra}. They are used as standard candles to measure galactic distances, assisting with the cosmic distance ladder. They are pulsating horizontal branch stars of spectral class A or F, with a mass of around half the Sun's. They are thought to have shed mass during the red-giant branch phase and were once stars of similar or slightly less mass than the Sun, around 0.8 solar masses. In contemporary astronomy, a period-luminosity relation makes them good standard candles for relatively nearby targets, especially within the Milky Way and Local Group. They are also frequent subjects in the studies of globular clusters. We use the set of photometric observations of RR Lyrae stars in the SDSS as our data. The data set comes from SDSS Stripe 82, and combines the Stripe 82 standard stars, which represent observations of non-variable stars; and the RR Lyrae variables pulled from the same observations as the standard stars, and selected based on their variability using supplemental data. The sample is further constrained to a smaller region of the overall color–color space following  $ 0.7<u - g<1.35, -0.15 < g - r < 0.4, -0.15 < r - i < 0.22, and -0.21 < i - z < 0.25$  \citep{Lyra}. These selection criteria lead to a sample of 92,658 non-variable stars, and 483 RR Lyraes. Two features of this combined data set make it a good candidate for testing classification algorithms:

\section{\large  Classification metrics}

When performing classification predictions, there are four types of outcomes that could occur. True positives, which are when you predict an observation belongs to a class and it actually does belong to that class. True negatives, which are when you predict an observation does not belong to a class and it actually does not belong to that class. False positives, which occur when you predict an observation belongs to a class when in reality it does not. False negatives, which occur when you predict an observation does not belong to a class when in fact it does. These four outcomes are often plotted on a confusion matrix. This matrix can be generated after making predictions on the test data and then identifying each prediction as one of the four possible outcomes described above.

The four main metrics used to evaluate a classification model are accuracy, precision, recall, and $F_1$ score.

Accuracy is defined as the percentage of correct predictions for the test data. It can be calculated easily by dividing the number of correct predictions by the number of total predictions.

\begin{equation}
\text{accuracy}=\frac{\text{correct predictions}}{\text{all predictions}}
\end{equation}

Precision is defined as the fraction of relevant examples (true positives) among all of the examples which were predicted to belong in a certain class.

\begin{equation}
\text{precision}=\frac{\text{true positives}}{\text{true positives+false positives}}
\end{equation}

Recall is defined as the fraction of examples which were predicted to belong to a class with respect to all of the examples that truly belong in the class.

\begin{equation}
\text{recall}=\frac{\text{true positives}}{\text{true positives+false negatives}}
\end{equation}

The $F_1$-score combines the precision and recall of a classifier into a single metric by taking their harmonic mean. It is primarily used to compare the performance of two classifiers. Suppose that classifier A has a higher recall, and classifier B has higher precision. In this case, the $F1$-scores for both the classifiers can be used to determine which one produces better results.

\begin{equation}
F_1=\frac{2 \text{ precision * recall}}{\text{precision + recall}}
\end{equation}

\section{\large Results}

\begin{figure}[th!]
\begin{center}
\includegraphics[width=0.95\textwidth]{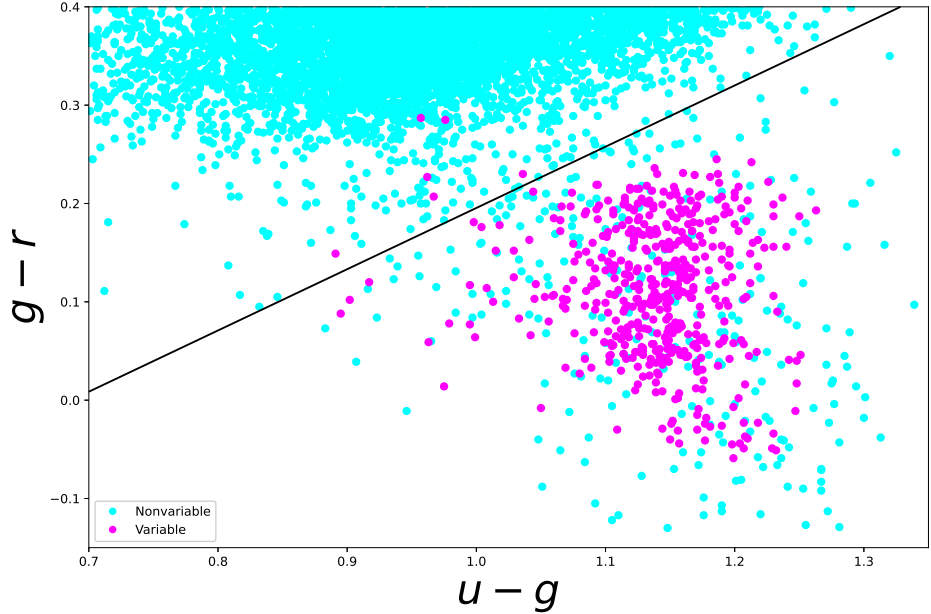}
\end{center}
\caption{Logistic Regression classification method is used to separate variable stars from nonvariable stars. The light blue points show nonvariable sources, while the violet show variable sources. The classification boundary is shown by the black line. Logistic Regression attains an accuracy of 0.969 and a $F_1$-score of 0.628.}
 \label{Fig1}
\end{figure}

\begin{figure}[th!]
\begin{center}
\includegraphics[width=0.95\textwidth]{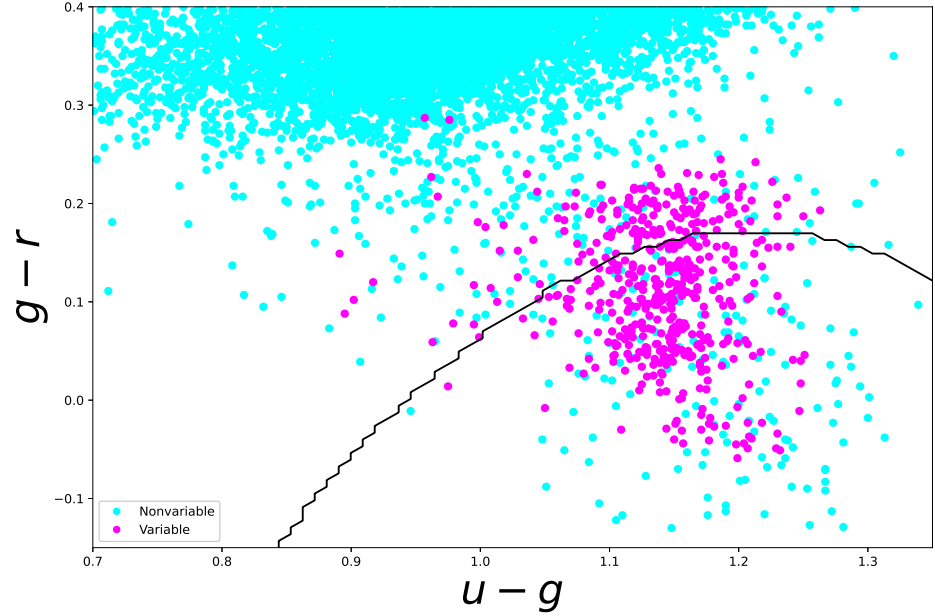}
\end{center}
\caption{Gaussian naive Bayes classification method is used to separate variable stars from nonvariable stars. The light blue points show nonvariable sources, while the violet show variable sources. The classification boundary is shown by the black line. Naive Bayes attains an accuracy of 0.983 and a $F_1$-score of 0.655.}
 \label{Fig2}
\end{figure}

In this section, we apply a collection of popular classifier on our data set to train a set of machine learning models and compare the different models with each other. In Figures (\ref{Fig1}, \ref{Fig2}), we have plotted our sample data and the classification boundary for a Logistic Regression and a Gaussian naive Bayes classifier.
In table (\ref{tabel1}) the accuracy and $F_1$-score for different models are reported which can be used to compare the proposed models.

\begin{table}
    \centering
    \begin{tabular}{l l l l }
        \hline
       Classifier &   Accuracy       &  $F_1 $                 \\
        \hline
        \vspace{0.3cm}
        Logistic regression  & 0.969 & 0.628 \\
        \vspace{0.3cm}
        Naive Bayesian   & 0.983    & 0.654    \\
        \vspace{0.3cm}  
        SVMs    & 0.994      & 0.499        \\
         \vspace{0.3cm}
        Decision Trees    & 0.994       & 0.703        \\
        \vspace{0.3cm}
        Neural Networks    & 0.993        & 0.512        \\
        
        \hline        
    \end{tabular}
    \caption{Accuracy and $F_1$ score for six chosen classifier. Although all models provide a very good accuracy, $F_1$ differs between proposed models and it is a better metric for evaluation of the models.}
    \label{tabel1}
\end{table}

\section{\large Summary and Conclusion}

We reviewed some of the popular classification algorithms including Linear Regression, Logistic Regression, Naive Bayesian, SVM, Decision Trees and Neural Networks. Then we applied some of the proposed models on a dataset of variable and nonvariable stars, extracted from SDSS survey Stripe 82. After training the models on the proposed dataset, we tested the models and evaluated their predictions. We used accuracy and $F_1$-score as our indicators, which are reported in Table (\ref{tabel1}) of the paper. 
As can be seen, the decision tree algorithm result in the best accuracy and $F_1$-score. In addition, for the sake of better understanding, we have plotted the decision boundary of a logistic regression model and a Naive Bayes model in Figures (\ref{Fig1}, \ref{Fig2})

\subsubsection*{Acknowledgements}

The author thanks the  organizers of ICRANet -- Isfahan Astronomy Meeting for their kind invitation to present a contribution for this activity.

\end{document}